\documentstyle[graphics]{mn}
\newif\ifAMStwofonts
\input psfig.sty

\def\source{A0538--66}
\def\til{$\sim$}
\def\deg{$^{\circ}$}
\def\ang{\thinspace\hbox{\AA}}
\def\tsp{\thinspace}

\title[A 421\tsp d Activity Cycle in A0538--66]
      {A 421\tsp d Activity Cycle in the BeX Recurrent Transient A0538--66 from
MACHO monitoring}
\author[C. Alcock et al.]
{C. Alcock$^{1,2}$, R. A. Allsman$^{3}$, D. R. Alves$^{4}$, 
T. S. Axelrod$^{5}$, A. C. Becker$^{6}$, 
\and
D. P. Bennett$^{7}$, P. A. Charles$^{8}$, K. H. Cook$^{1,2,9}$, 
A. J. Drake$^{1}$, K. C. Freeman$^{5}$, 
\and
M. Geha$^{10}$, K. Griest$^{2,11}$, M. J. Lehner$^{12}$, 
S. L. Marshall$^{1,9}$, K. E. McGowan$^{8}$\thanks{email: kem@astro.ox.ac.uk},
\and
D. Minniti$^{1,13}$, C. A. Nelson$^{14}$, B. A. Peterson$^{5}$, 
P. Popowski$^{1}$, M. R. Pratt$^{15}$,
\and
P. J. Quinn$^{16}$, C. W. Stubbs$^{2,6}$, W. Sutherland$^{8}$, 
A. B. Tomaney$^{6}$, T. Vandehei$^{11}$, 
\and
D. L. Welch$^{17}$\\
$^{1}$Lawrence Livermore National Laboratory, Livermore, CA 94550, USA\\
$^{2}$Center for Particle Astrophysics, University of California, Berkeley, CA 
94720, USA\\
$^{3}$Supercomputing Facility, Australian National University, Canberra, ACT 
0200, Australia\\
$^{4}$Space Telescope Science Institute, Baltimore, MD 21218, USA\\
$^{5}$Research School of Astronomy and Astrophysics, Canberra, Weston Creek, 
ACT 2611, Australia\\
$^{6}$Departments of Astronomy and Physics, University of Washington, Seattle, 
WA 98195, USA\\
$^{7}$Department of Physics, University of Notre Dame, Notre Dame, IN 46556, 
USA\\
$^{8}$Department of Astrophysics, Nuclear Physics Laboratory, Keble Road, 
Oxford OX1 3RH\\
$^{9}$Visiting Astronomer, Cerro Tololo Inter-American Observatory, which is 
operated by the Association of Universities for Research in\\
Astronomy, Inc., under cooperative agreement with the National Science 
Foundation\\
$^{10}$Department of Astronomy and Astrophysics, UC Santa Cruz, Santa Cruz, CA
90064\\
$^{11}$Department of Physics, University of California, San Diego, La Jolla, CA
92093, USA\\
$^{12}$Department of Physics, University of Sheffield, Sheffield, S3 7RH\\
$^{13}$Departmento de Astronomia, P. Universidad Catolica, Casilla 104, 
Santiago 22, Chile\\
$^{14}$Department of Physics, University of California, Berkeley, CA 94720, 
USA\\
$^{15}$Center for Space Research, MIT, Cambridge, MA 02139, USA\\
$^{16}$European Southern Observatory, Karl-Schwarzchild Strasse 2, D-85748, 
Garching, Germany\\
$^{17}$Department of Physics and Astronomy, McMaster University, Hamilton, 
Ontario, L8S 4M1, Canada}

\pagerange{\pageref{firstpage}--\pageref{lastpage}}
\pubyear{2000}

\begin{document}

\maketitle

\label{firstpage}

\begin{abstract}
{We present a \til5-yr optical light curve of the recurrent Be/X-ray transient 
\source\ obtained as a by-product of the MACHO Project.  These data reveal both
a long-term modulation at P = 420.8$\pm$0.8 d and a short-term modulation 
at 16.6510$\pm$0.0022 d which, within errors, confirms the previously 
found orbital period.  Furthermore, the orbital activity is only seen at 
certain phases of the 421 d cycle suggesting that the long-term modulation 
is related to variations in the Be star envelope.}

\end{abstract}


\section{Introduction}

The recurrent X-ray transient \source\ was discovered with {\it Ariel V} when 
two outbursts, separated by \til17 d, were observed (White and Carpenter 1978).
Further outbursts were observed with {\it HEAO-1} which, when the source was 
active, was found to have a periodicity of 16.668 d (Johnston et al.\ 1979; 
Johnston et al.\ 1980; Skinner et al.\ 1980; Skinner 1980), the precision of 
which led to its interpretation as being orbital.  Skinner (1981) used archival
plates taken over \til50 years in order to obtain an improved value of 16.6515 
d for this periodicity, based on the recurrence of the outbursts.

\source\ has been optically identified with a B star (V\til15) (Johnston et 
al.\ 1980), and an X-ray pulse period of 69 ms indicates that the compact 
object is a neutron star (Skinner et al.\ 1982).  In quiescence (X-ray 'off') 
the colour, magnitude and radial velocity are consistent with a B2 III-IV star 
in the LMC (Charles et al.\ 1983), but during outburst (which can reach as high
as V\til12; see Densham et al.\ 1983) it is redder with a spectral type of 
B8-9 I.  During long phases of inactivity (there have been no large X-ray or 
optical outbursts reported since 1983) there is evidence that a remnant 
envelope is still present in the system (Smale et al.\ 1984).  \source\ has 
usually been interpreted as a Be-type system which exhibits extreme outbursts 
as a result of its neutron star companion interacting with the Be star in a 
highly eccentric orbit (Charles et al.\ 1983).

Here we exploit the fortuitous location of \source\ in a MACHO field and the 
extended (\til5 yrs), regular (\til nightly) monitoring by the MACHO project 
in order to examine the quiescent variability of this, the most X-ray luminous 
of all Be X-ray transients.

\section{Observations}

The MACHO observations were made using the 1.27 m telescope at Mount Stromlo 
Observatory, Australia.  A dichroic beamsplitter and filters provide 
simultaneous CCD photometry in two passbands, a 'red' band (\til6300--7600 
\ang) and a 'blue' band (\til4500--6300 \ang).  The latter filter is a broader
version of the Johnson {\it V} passband (see Alcock et al.\ 1995a, 1999 for 
further details).

The images were reduced with the standard MACHO photometry code {\scshape 
sodophot}, based on point-spread function fitting and differential photometry 
relative to bright neighbouring stars.  Further details of the instrumental 
set-up and data processing may be found in Alcock et al.\ (1995b, 1999), 
Marshall et al. (1994) and Stubbs et al.\ (1993).

\section{MACHO Project Photometry}

\subsection{Light curve}

We show in Fig. 1 the 'blue' and 'red' photometry of \source, the magnitudes of
which were transformed to Johnson {\it V} and Kron-Cousins {\it R} 
respectively, using the absolute calibration of the MACHO fields.  This 
absolute calibration depends on the colour of the object itself; therefore, at 
times when there was no red data available a mean colour for the data was used.
The data consist of MACHO project observations taken during the period 1993 
January 14 to 1998 May 28.  There are fewer data points in the {\it R}-band 
because half of one of the four CCD chips in the red focal plane is
inoperative.  Due to the German telescope mount, the focal plane may be rotated
by 0\deg\ or 180\deg\ relative to the sky (depending on hour angle), hence 
\source\ falls on the dead area in about half the observations.  As the 
{\it V}-band light curve has denser sampling than the {\it R}-band, only the 
{\it V}-band light curve was used in the rest of the analysis. 

\subsection{Period analysis and folded light curve}

To search for periodicities in the {\it V}-band light curve two different 
frequency domain techniques were employed : (i) we calculated a Lomb-Scargle 
(LS) periodogram (Lomb 1976; Scargle 1982) on the dataset, to search for 
sinusoidal modulations [this periodogram is a modified discrete Fourier 
transform (DFT), with normalizations which are explicitly constructed for the 
general case of time sampling, including uneven sampling; see Scargle 1982]; 
(ii) we constructed a phase dispersion minimisation (PDM) periodogram, which 
works well even for highly non-sinusoidal light curves (see Stellingwerf 1978).

As a modulation on longer timescales is clearly evident in the light curve the 
data were searched over the frequency range 0.001-0.01 cycle d$^{-1}$ with a 
resolution of 1x$10^{-6}$ cycle d$^{-1}$ (Figure 2).  To search for modulations
in the region of the previously quoted 16.6 d period the long term variations 
were removed.  The data were split into 15 sections and each section was 
detrended separately by subtracting a linear fit.  A frequency space of 
0.01-1.2 cycle d$^{-1}$ was searched with a resolution of 1x$10^{-4}$ cycle 
d$^{-1}$ (Figure 5).

\subsubsection{The 421 d period}

In the longer period search a dominant peak is found in the LS periodogram at 
P = 420.52 d (Figure 2).  The peak in the LS has a confidence of greater 
than 99\% as determined from a cumulative probability distribution (CDF) 
appropriate for the data set.  By constructing the cumulative probability 
distribution (CDF) of the random variable $P_{\it X}$($\omega$), the power at 
a given frequency, where {\it X} is pure noise (Scargle 1982), we can measure 
the significance of the peaks in the LS periodogram.  In practice the CDF was 
constructed using a Monte Carlo simulation method.  Noise sets with the same 
sampling as the MACHO data were generated, and the LS periodogram was run upon 
each one.  The peak power occurring in the periodogram due purely to noise was 
then recorded.  This was repeated for ten-thousand noise sets, to produce good 
statistics.  From these values the probability of obtaining a given peak power 
from pure noise can then be calculated and the CDF derived.  In order to test 
the significance of peaks from a given data set, the generated noise sets 
should have the same variance.  Therefore, each noise set was generated from a 
random number generator that takes values from a Gaussian distribution with 
the same variance as the data set.  The dip in the PDM periodogram 
corresponding to the peak in the LS periodogram is broad and highly structured,
therefore a centroiding technique was employed to calculate the mode of the 
dip giving a period of 420.82 d (Figure 2).

The MACHO {\it V}-band data were folded on P = 420.82 d (Figure 3, top panel), 
and then binned (Figure 3, middle panel), to examine the form of the modulated 
variability.  Burst points that correspond to the 16.6 d period are evident in 
the data (see Section 3.2.2), these were removed and the remaining data was 
bin-folded (Figure 3, bottom panel).  To see if the broadness of the peak 
(FWHM = 6.0x$10^{-4}$ d$^{-1}$) in the LS periodogram for the longer period 
search was due to it being quasi-periodic we simulated a truly sinusoidal light
curve to be used in period searching.  The light curve was produced using a 
Gaussian random number generator with the same mean and standard deviation as 
the data plus a sinusoid with a frequency set to the period found.  A period 
search was performed over the frequency range 0.001-0.01 cycle d$^{-1}$ with 
1x$10^{-6}$ cycle d$^{-1}$ resolution.  The resulting LS periodogram is shown 
as a dashed line in Figure 4.  The peak produced from the simulated light curve
has a FWHM = 5.5x$10^{-4}$ d$^{-1}$ and is almost as broad as that for the real
data which shows that the modulation found is truly periodic.

In order to estimate the uncertainty in the 420.82 d dip in the PDM periodogram
we performed a Monte Carlo simulation in which we created artificial light 
curves with the same mean and standard deviation as the {\it V}-band data.  
Phase dispersion minimisation periodograms were constructed and the minimum 
value found for each artificial dataset using the centroiding technique near 
the dip of interest was recorded.  The results had an average scatter of 
$\pm$0.79 d.

\subsubsection{The 16.6 d period}

The resulting LS periodogram for the short period search has a peak with much
greater than 99\% confidence at P = 16.6667 d, together with two marginally 
significant peaks around 1 d (Figure 5).  To identify the origin of these peaks
more simulated datasets were created, which had the same mean and standard 
deviation as the detrended data.  These were produced using a Gaussian random 
number generator plus a sinusoid with period 16.6667 d.  The resulting light 
curve was searched over a frequency range of 0.01-1.2 cycle d$^{-1}$ with a 
resolution of 1x$10^{-4}$ cycle d$^{-1}$.  As can be seen from the resulting LS
periodogram (Figure 5) the two peaks are 1 d aliases produced due to the 
sampling of the dataset.  Using the centroiding technique again the dip in the 
PDM periodogram for the short period search was found to correspond to P = 
16.6510 d (Figure 5).  

The detrended {\it V}-band data were folded on P = 16.6667 d and P = 16.6510 d 
using Skinner's ephemeris (1982).  The folded light curves showed that the 
modulation was highly non-sinusoidal, hence we took the PDM value for the 
orbital period, and the bin-folded light curve using P = 16.6510 d is shown in 
Figure 6.  The error on the period was propagated using the same method as for 
the long period search producing a value of $\pm$0.0022 d.

\section{Discussion}

These extensive MACHO observations have revealed, not only a remarkably stable 
long-term modulation, but also the presence of the same 16.6 d orbital 
modulation (in the form of mini-outbursts) that had been seen (on a larger 
scale) in the 1980's.  But the key additional point is that these 
mini-outbursts are constrained in 421 d phase to only occur during minima, 
thereby establishing a physical link between the two.  What can this be?

To explain the origins of the long-term variability in \source\ it is 
instructive to consider other systems that show modulations on these 
timescales.  The soft X-ray transient (SXT) 4U\tsp1630-47 is the shortest known
recurrent black-hole SXT, with an outburst recurrence interval of \til600-690 d
(Jones et al.\ 1976, Priedhorsky 1986, Kuulkers et al.\ 1997).  The outburst 
recurrence times vary, as do the intensities at the peak of the outburst and 
its duration (Kuulkers et al.\ 1997), this behaviour is also seen in \source\ 
(Skinner et al. 1980).  Both sources also undergo transitions between high and 
low activity on timescales \til1 yr (4U\tsp1630-47; Kuulkers et al.\ 1997, 
\source; Skinner et al.\ 1980, Pakull \& Parmar 1981), and exhibit 
inter-outburst activity (4U\tsp1630-47; Parmar et al.\ 1997, Kuulkers et al.\ 
1997, \source; Densham et al.\ 1983).  Kuulkers et al. (1997) showed that the 
outburst recurrence time cannot be related to orbital variations of 
4U\tsp1630-47.  Due to heavy reddening the optical counterpart has not yet been
detected in 4U\tsp1630-47, and so a Be-type secondary cannot be excluded.

Analysis of X-ray observations of another Be/X-ray transient, A0535+26, led to 
an orbital period of 110.3 d for the source (Finger et al.\ 1996).  Analysis of
the long term optical light curve by Clark et al.\ (1999) showed periodicities 
at \til1400 d, \til476 d and \til103 d, but no evidence was found for the 
\til110 d presumed orbital period of the neutron star.  Clark et al.\ (1999) 
could not identify the origin of these quasi-periods, and it was not clear if 
they were coherent over time.  

In normal Be star shell ejection events, which recur on timescales of years, 
the circumstellar matter lost forms an equatorial disc around the Be star due 
to its rapid rotation (Slettebak 1987).  The Paschen continuum of the 
equatorial envelope is in emission and thus emits optical light which is redder
in {\it B-V} than that emitted by the early type star.  The formation of the 
disc will either add red light and increase the optical brightness of the 
system or will mask the early type star and make it appear fainter, depending 
on the inclination (Corbet et al.\ 1986, Janot-Pacheco et al.\ 1987).  The 
photometric behaviour of the Be star in \source\ in the {\it V}, 
{\it V-R} diagram (Figure 7) shows a reddening as the source gets fainter.
The fading and rising events in \source's light curve could be due to the 
formation and depletion of an equatorial disc seen at high inclination.

If this is the case, the fact that the neutron star is believed to be in an 
eccentric orbit around the Be star could set a limit on the size of the 
equatorial disc that can form producing the 421 d periodicity.  Using this 
model the mini-outbursts only occur when the disc reaches the upper limit set 
by the neutron star's orbit.  Although Okazaki (1998) and Negueruela (1999) 
suggest that the presence of the neutron star in orbit around the Be star will 
tidally truncate the circumstellar disc of the Be star well within the orbital 
radius of the neutron star, thus preventing accretion, modelling by Haynes et 
al.\ (1980) and Brown and Boyle (1984) shows that matter can escape from the 
Be star and be spread over a domain in the vicinity of the orbit of the compact
star near periastron.  Therefore the 'flat' part of \source's light curve is 
when we are seeing the normal B star without any equatorial disc, hence no 
outbursts can occur.  If this is correct, the previously obtained spectra 
thought to be taken during 'quiescence' were really taken during the extended 
dips in \source's light curve.  This indicates that the true quiescent 
magnitude of the system is \til V=14.4, and the quoted quiescent 
spectral-type must be re-determined.  As the colours obtained from the MACHO 
data are not precise enough we cannot confirm the previously determined 
spectral classification for \source.  A study of archival data of \source, 
including data taken before 1980, will be presented in a future publication.

Skinner (1981) found the recurrence of the outbursts of \source\ to be 16.6515
$\pm$0.0005 d or 16.6685$\pm$0.0005 d.  We find the periodicity to be
16.6510$\pm$0.0022 d, which within errors confirms the first of Skinner's 
previously suggested orbital periods for \source.

\section{Acknowledgements}

KEM acknowledges the support of a PPARC studentship.  KEM and PAC thank Malcolm
Coe for useful discussions.
DM is supported by Fondecyt 1990440.  This work was performed under the 
auspices of the U.S. Department of Energy by University of California Lawrence 
Livermore National Laboratory under contract No. W-7405-Eng-48.

\newpage

\begin{figure*}
\resizebox{1.0\textwidth}{!}{\rotatebox{-90}{\includegraphics{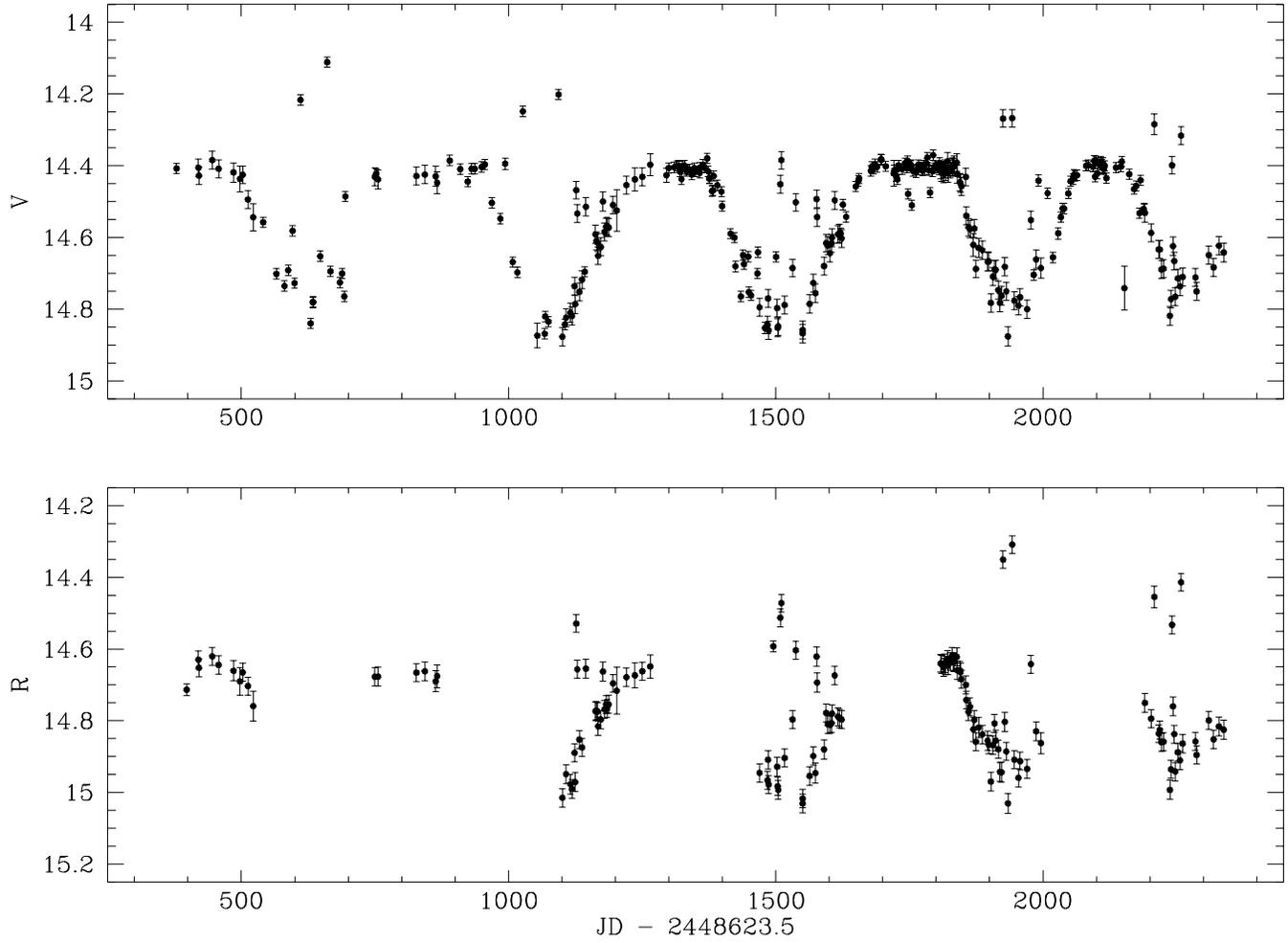}}}
\caption{The {\it V}- ({\it top panel}) and {\it R}-band ({\it bottom panel}) 
light curves of \source\ from MACHO project observations.  The {\it V} and 
{\it R} magnitudes have been calculated using the absolute calibrations of the 
MACHO fields.}
\end{figure*}

\begin{figure*}
\resizebox{1.0\textwidth}{!}{\rotatebox{-90}{\includegraphics{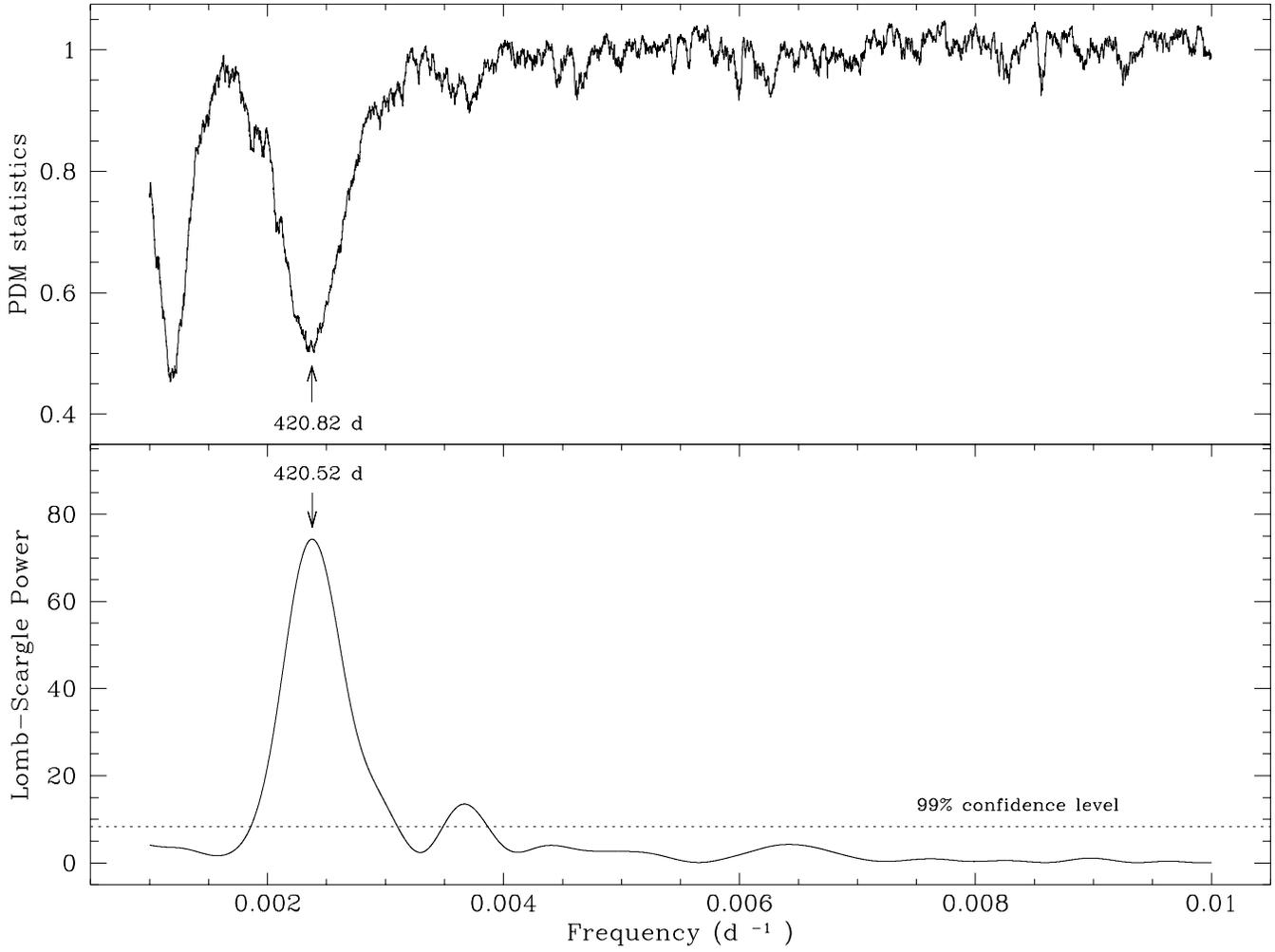}}}
\caption{Phase dispersion minimisation periodogram ({\it top panel}) and 
Lomb-Scargle periodogram ({\it bottom panel}) for long period search of 
{\it V}-band data, frequency space 0.001-0.01 cycle d$^{-1}$ and resolution 
1x$10^{-6}$ cycle d$^{-1}$.}
\end{figure*}

\begin{figure*}
\resizebox{1.0\textwidth}{!}{\rotatebox{-90}{\includegraphics{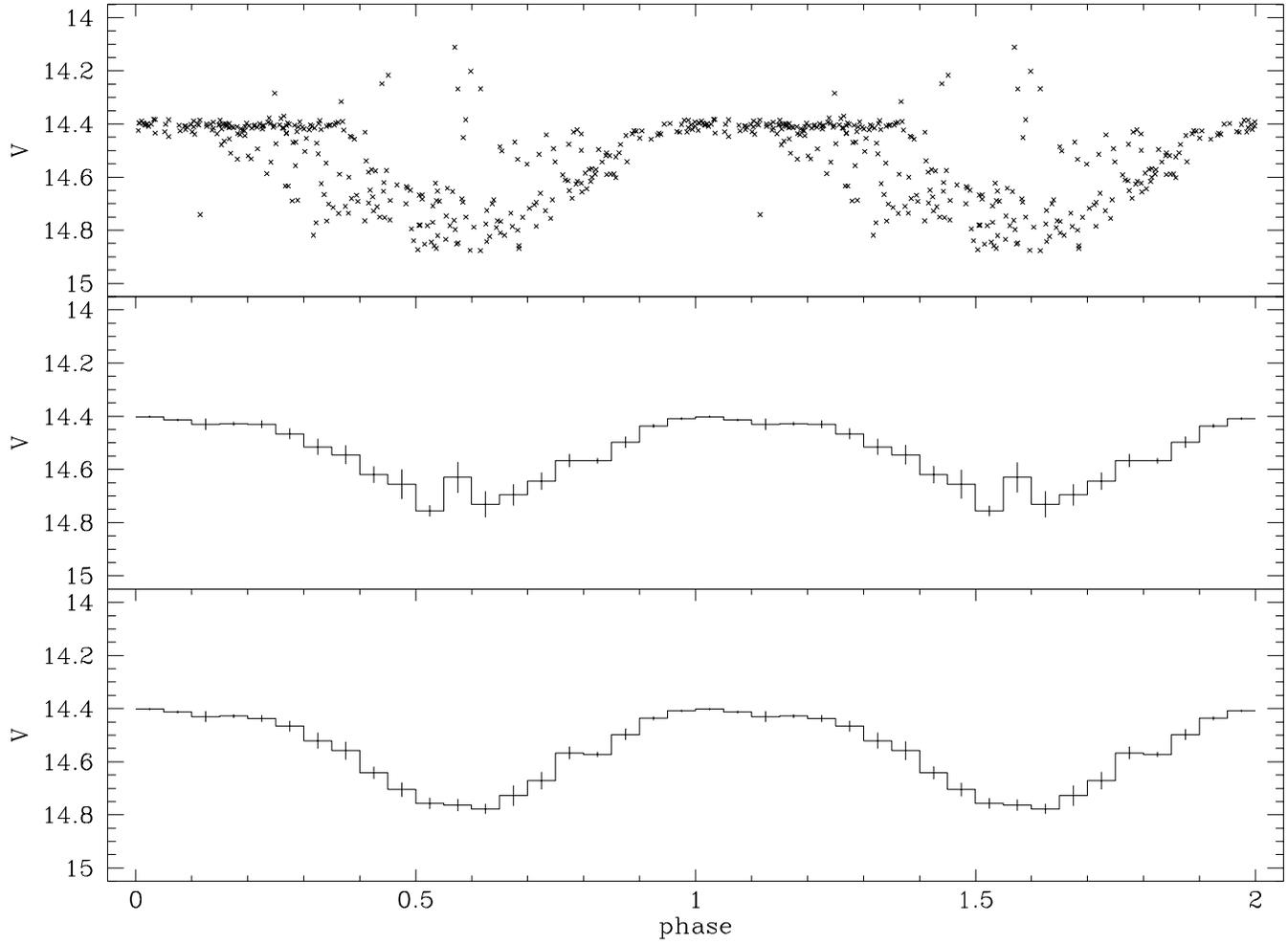}}}
\caption{{\it V}-band data folded on P = 420.82 d using T$_{\circ}$ = JD 
2449002.109 ({\it top panel}), and in 20 phase bins ({\it middle panel}).  In 
the {\it bottom panel} the 16.6510 d burst points have been removed from the 
data which was then folded into 20 phase bins.  Error bars for both binned 
light curves are the standard errors for the data points in each bin.}
\end{figure*}

\begin{figure*}
\resizebox{1.0\textwidth}{!}{\rotatebox{-90}{\includegraphics{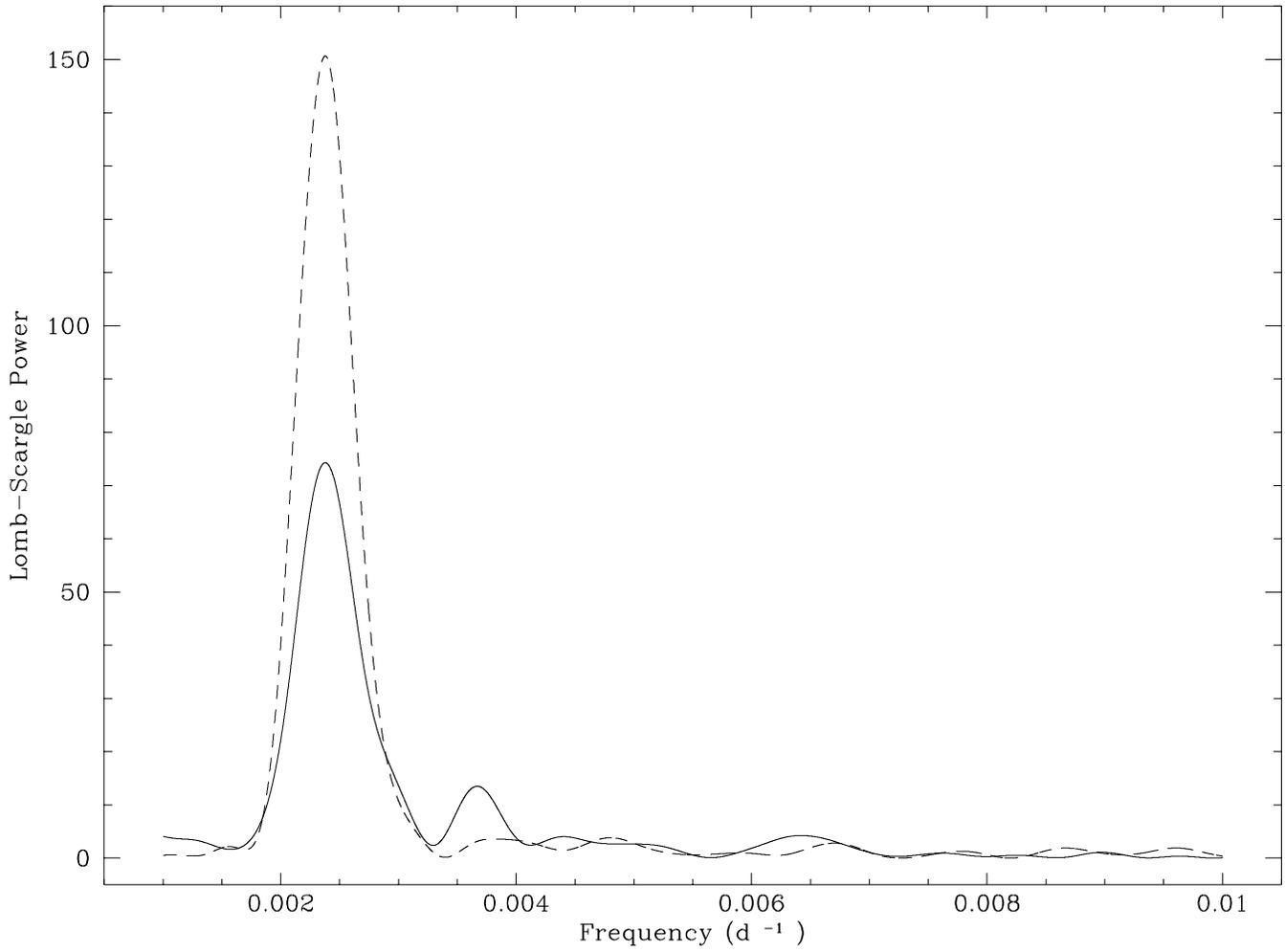}}}
\caption{Lomb-Scargle periodogram for long period search of {\it V}-band data 
(solid line) and simulated data (dashed line), frequency space 0.001-0.01 
cycle d$^{-1}$ and resolution 1x$10^{-6}$ cycle d$^{-1}$.}
\end{figure*}

\begin{figure*}
\resizebox{1.0\textwidth}{!}{\rotatebox{-90}{\includegraphics{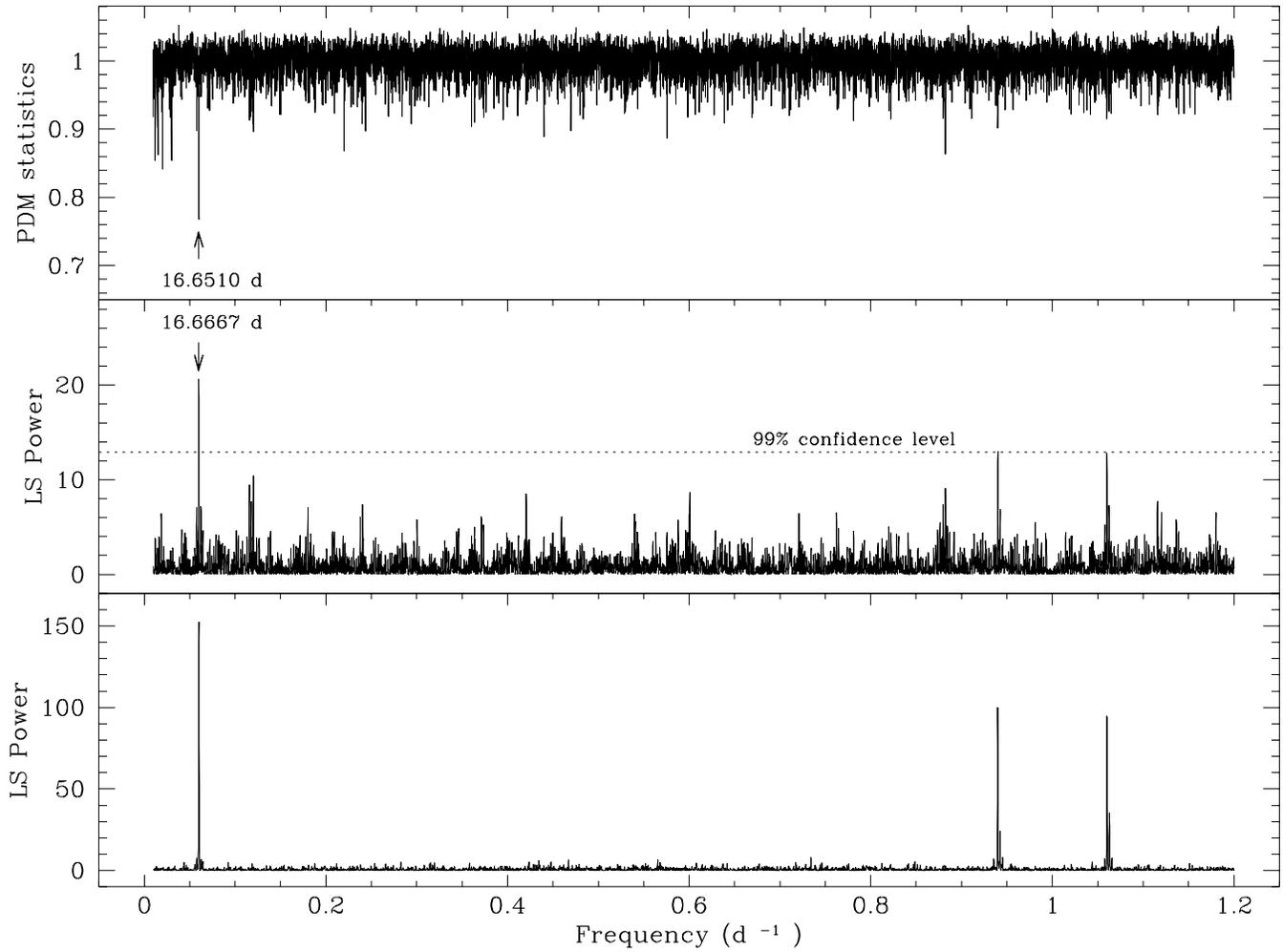}}}
\caption{Phase dispersion minimisation periodogram ({\it top panel}) and 
Lomb-Scargle periodogram ({\it middle panel}) for short period search of 
detrended {\it V}-band data, frequency space 0.01-1.2 cycle d$^{-1}$ and 
resolution 1x$10^{-4}$ cycle d$^{-1}$.  {\it Bottom panel}, Lomb-Scargle 
periodogram for the simulated light curve which contained a sinusoidal signal 
of 16.6667 d.}
\end{figure*}

\begin{figure*}
\resizebox{1.0\textwidth}{!}{\rotatebox{-90}{\includegraphics{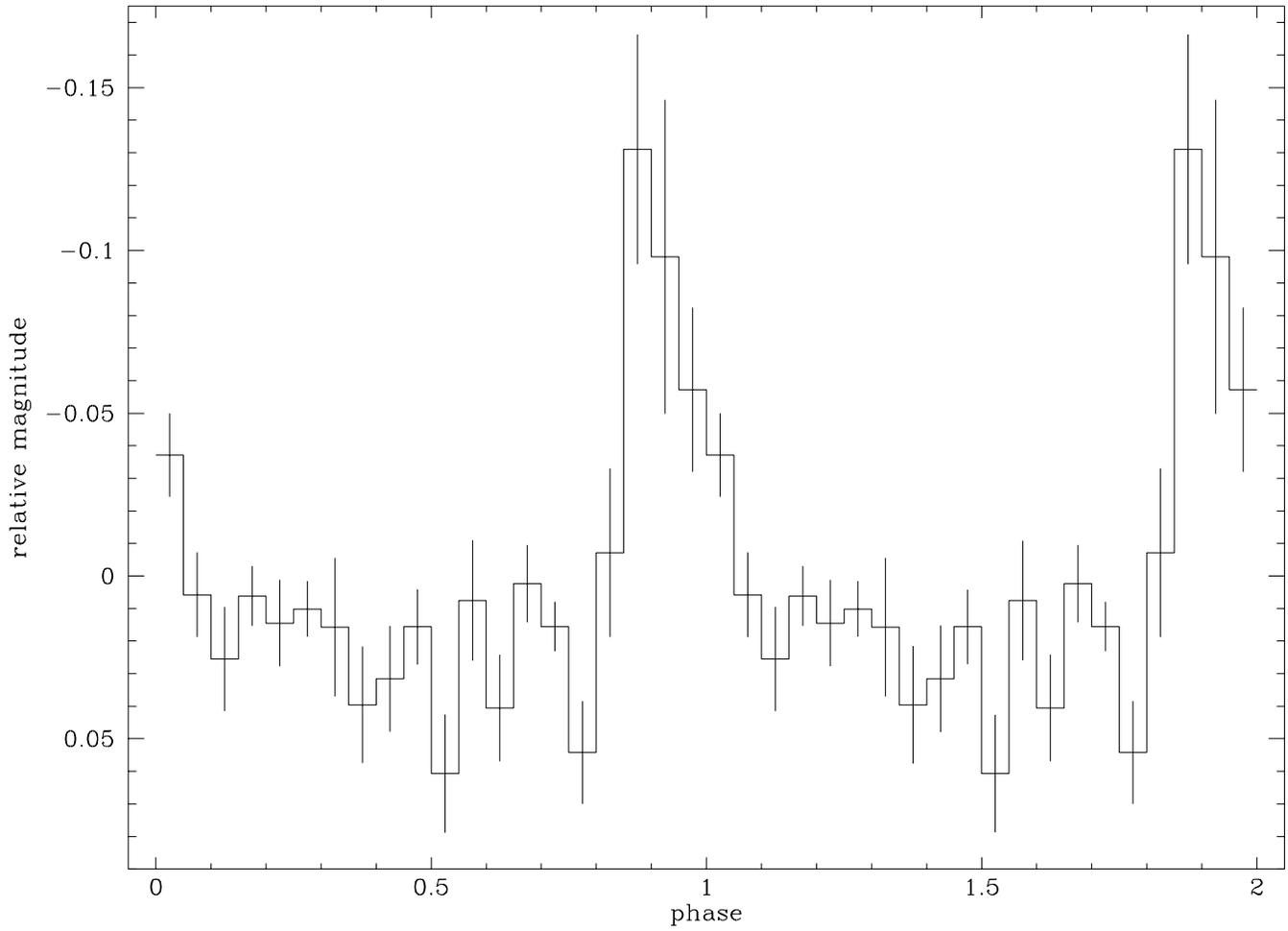}}}
\caption{{\it V}-band data, detrended and folded on P = 16.6510 d in 20 phase 
bins, using T$_{\circ}$= MJD 2443423.96$\pm$0.05 (Skinner 1981).  Error 
bars for the binned light curve are the standard errors for the data points in 
each bin.}
\end{figure*}

\begin{figure*}
\resizebox{1.0\textwidth}{!}{\rotatebox{-90}{\includegraphics{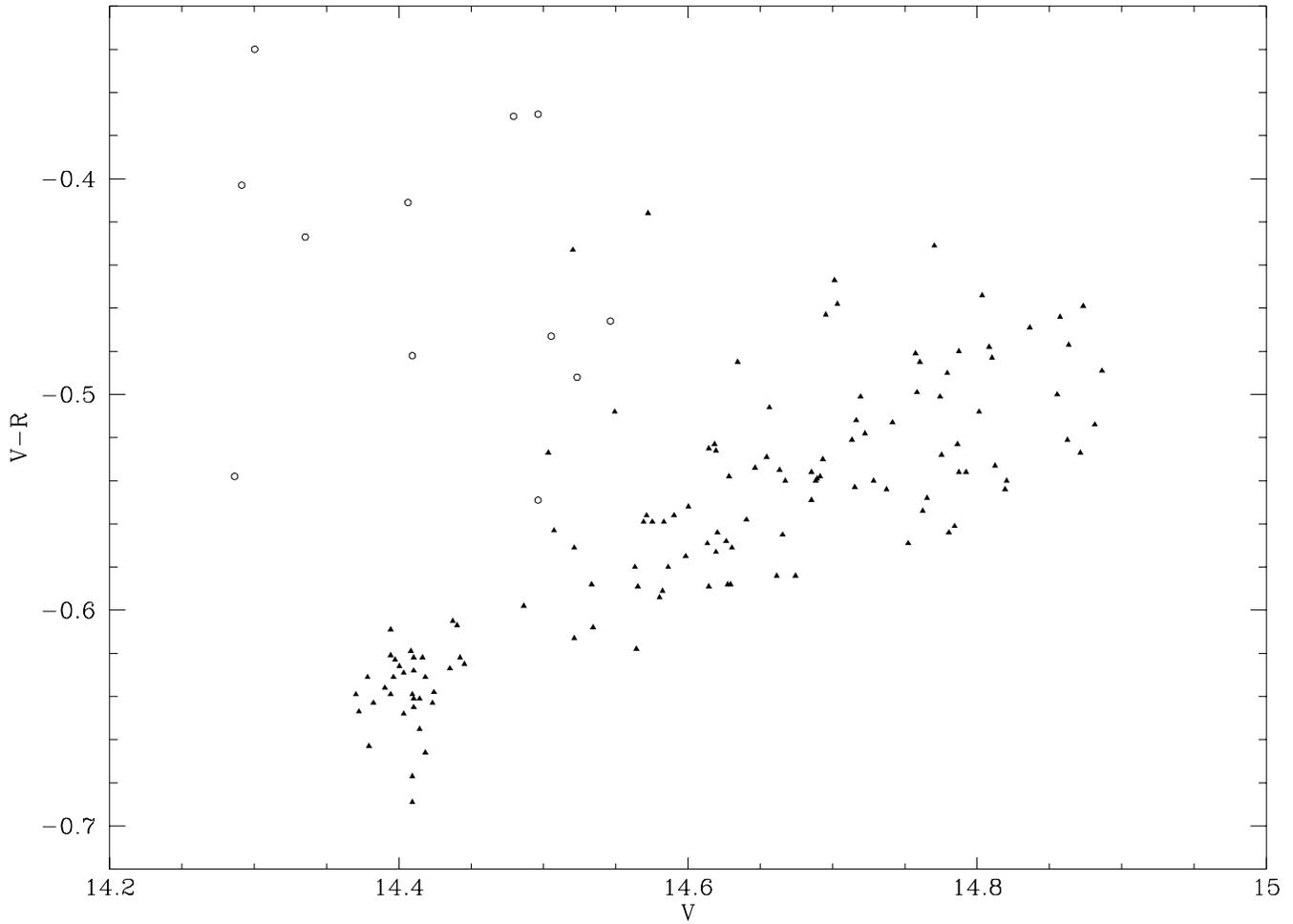}}}
\caption{Colour magnitude diagram of \source\ showing that the star is redder 
when fainter which is interpreted in classical Be models as a consequence of 
the formation of an equatorial envelope seen at high inclination.  Open circles
correspond to the 16.6510 d burst points, filled triangles to all other data.  
The {\it V-R} values are purely instrumental , whereas the {\it V} magnitudes 
have been calculated using the absolute calibrations of the MACHO fields (see 
Section 3.1 for details).}
\end{figure*}

\label{lastpage}

\end{document}